\documentclass[conference]{IEEEtran}

\usepackage{graphicx}
\usepackage{multirow}
\usepackage{tabularx}
\usepackage{array}
\usepackage[hyphens]{url}
\usepackage{hyperref}
\PassOptionsToPackage{hyphens}{url}
\renewcommand\IEEEkeywordsname{Keywords}

\newcolumntype{Y}{>{\centering\arraybackslash}X}

\begin{document}
\title{You Do Not Need More Data: Improving End-To-End Speech Recognition by Text-To-Speech Data Augmentation}

\author{\IEEEauthorblockN{Aleksandr Laptev\IEEEauthorrefmark{1}, Roman Korostik\IEEEauthorrefmark{1}, Aleksey Svischev\IEEEauthorrefmark{1}, Andrei Andrusenko\IEEEauthorrefmark{1},\\ Ivan Medennikov\IEEEauthorrefmark{1}$^{,}$\IEEEauthorrefmark{2}, Sergey Rybin\IEEEauthorrefmark{1}$^{,}$\IEEEauthorrefmark{2}}
\IEEEauthorblockA{\IEEEauthorrefmark{1}ITMO University,
St. Petersburg, Russia 197101}
\IEEEauthorblockA{\IEEEauthorrefmark{2}STC-innovations Ltd,
St. Petersburg, Russia 194044\\
Email: \{laptev, korostik, svishchev, andrusenko, medennikov, rybin\}@speechpro.com}}


%Atlanta, Georgia 30332--0250}
%\IEEEauthorblockA{\IEEEauthorrefmark{2}Twentieth Century Fox, Springfield, USA}
%Telephone: (800) 555--1212, Fax: (888) 555--1212}


\maketitle

\begin{abstract}
%\boldmath
Data augmentation is one of the most effective ways to make end-to-end automatic speech recognition (ASR) perform close to the conventional hybrid approach, especially when dealing with low-resource tasks. Using recent advances in speech synthesis (text-to-speech, or TTS), we build our TTS system on an ASR training database and then extend the data with synthesized speech to train a recognition model. We argue that, when the training data amount is relatively low, this approach can allow an end-to-end model to reach hybrid systems' quality. For an artificial low-to-medium-resource setup, we compare the proposed augmentation with the semi-supervised learning technique. We also investigate the influence of vocoder usage on final ASR performance by comparing Griffin-Lim algorithm with our modified LPCNet. When applied with an external language model, our approach outperforms a semi-supervised setup for LibriSpeech \textit{test-clean} and only 33\% worse than a comparable supervised setup. Our system establishes a competitive result for end-to-end ASR trained on LibriSpeech \textit{train-clean-100} set with WER 4.3\% for \textit{test-clean} and 13.5\% for \textit{test-other}.
\end{abstract}


% \fi
\IEEEpeerreviewmaketitle

\section{Introduction}

There are two main approaches to automatic speech recognition (ASR): hybrid (deep neural networks combined with Hidden Markov models (DNN-HMM)) and end-to-end (jointly trained neural network systems). When training data amount is sufficiently large, they perform nearly equal \cite{luscher_rwth_2019}. But for low-resource tasks, end-to-end recognition quality is far behind from the hybrid models \cite{Andrusenko2020TowardsAC,Medennikov_chime6}.

However, data augmentation techniques can help to make the end-to-end approach more competitive in the low-resource case \cite{Wiesner2019}. One can augment the data by morphing the training set itself or by obtaining additional data. Semi-supervised learning is one of the external data usage techniques proven to be effective \cite{synnaeve2019endtoend}. It consists of using a union of labeled training data and unlabeled data with transcription produced by the initial model to train a new one. However, it might not get the desired improvement if the initial model's recognition quality is low. There are other approaches, such as transfer and active learning \cite{drugman2016active}, to tackle low-data challenges, but they are beyond the scope of this paper.

Recent advances in speech synthesis (text-to-speech or TTS) made synthesizing close-to-human speech possible \cite{shen2018natural}; moreover, recent architectures allow unsupervised modeling of prosody, which allows to synthesize same texts with diverse speaking styles. Apart from human-computer interaction applications, such high-quality artificial speech can be used as a data augmentation for ASR \cite{li2018training}. It can be done either by training a standalone multi-speaker TTS system on in-domain data \cite{rosenberg_speech_2019} or by building and training of ASR and TTS systems jointly \cite{tjandra_end--end_2019}. It also has been shown that widely used augmentation techniques, specifically speed perturbation \cite{Ko_2015} and SpecAugment \cite{Park_2019}, stack effectively with TTS augmentation \cite{rossenbach_generating_2020}.

This work aims to improve speech recognition quality by augmenting training data using speech synthesis. Conventional low-resource tasks contain tens of hours (or just several hours) of transcribed data and no external text data. In contrast, in this work we follow \cite{rossenbach_generating_2020} and use LibriSpeech database \cite{panayotov_librispeech:_2015-1} to simulate a so-called low-to-medium-resource (here stands for one hundred hours) task. First, our TTS system and baseline ASR system are trained separately on the same 100-hour subset of training data. Then we synthesize utterances from a larger subset and use them as training data for ASR. We compare this approach with a semi-supervised learning technique to determine which one is more suitable for use in the setup in question. We also investigate the influence of waveform generation methods in TTS on final ASR performance by using the Griffin-Lim algorithm \cite{griffin_speech_1984} and our modification of LPCNet \cite{valin2019lpcnet}. Finally, we present our findings on how the proposed TTS augmentation stacks with the use of a language model and compare our setup with prior works.

\section{Related work}
\label{rel_work}

There are four prior works similar to ours. All of them use LibriSpeech to train their end-to-end ASR models and employ style modeling in TTS systems.

Li et al. \cite{li2018training} is the earliest of these works. Authors used Global Style Token (GST) \cite{yuxuan_style_2018} to model prosody. Unlike later works, authors trained their TTS model on an external three-speaker dataset (M-AILABS\footnote{\url{https://www.caito.de/2019/01/the-m-ailabs-speech-dataset/}}). Authors used Tacotron 2 as synthesizer, WaveNet as vocoder, and Wave2Letter+ ASR model from OpenSeq2Seq toolkit\footnote{\url{https://github.com/NVIDIA/OpenSeq2Seq}}.

Rosenberg et al. \cite{rosenberg_speech_2019} used a hierarchical version of variational autoencoder (VAE) \cite{kingma2013auto} to model prosody. They also addressed the lexical diversity problem (the number of real transcriptions is limited) by generating new utterances using a language model. The authors used their own Tacotron 2, WaveRNN, and an attention-based ASR model.

Sun et al. \cite{sun2020generating} suggest modeling style using an autoregressive prior over representations from quantized fine-grained VAE and perform evaluation by synthesizing utterances for training ASR on LibriSpeech. Authors do not consider a low- or medium-resource setup and do not report details of their ASR system.

Rossenbach et al. \cite{rossenbach_generating_2020} work is the most similar to our research in terms of the experimental setup. It compares GST with i-vectors for style modeling task. The authors used Tacotron 2, Griffin-Lim algorithm, and an attention-based ASR model from RETURNN framework\footnote{\url{https://github.com/rwth-i6/returnn}}.

\section{ASR system}

\subsection{Acoustic model}

As an ASR model, we chose the Transformer from the ESPnet\footnote{\url{https://github.com/espnet/espnet}} speech recognition toolkit \cite{watanabe_espnet:_2018}, since it delivers close to state-of-the-art results on LibriSpeech \cite{karita_comparative_2019}. Transformer \cite{attention_2017} is a sequence-to-sequence (S2S) architecture that uses self-attention mechanism to employ sequential information. It learns to transform one (source) sequence to another (target). Transformer consists of two neural networks: the Encoder and the Decoder. The Encoder transforms a source sequence into an intermediate sequence. This sequence is used for connectionist temporal classification (CTC) \cite{graves_connectionist_2006} frame-wise posterior distribution computing. The Decoder network also uses this intermediate representation along with previous target frames to predict the next target frame distribution. The final prediction is made using beam search and is computed as a weighted sum of CTC and S2S decoding posteriors.

\subsubsection{Encoder}

The Encoder network is a sequential module with two subnetworks and a positional encoding submodule between them. The first network transforms and subsamples an input acoustic feature sequence by using a two-layer CNN with 256 units, kernel size 3, and stride 2, yielding a four times shorter sequence. This subsampled sequence is added to a sinusoidal positional encoding tensor, which maps the position of each feature unit for each timestamp to the corresponding number, thereby allowing the Transformer to operate with the order of the sequence. The second network encodes the sequence by multiple sequential structures of one multi-head self-attention (MHA) layer and two feedforward (FF) layers, according to \cite{attention_2017}.

\subsubsection{Decoder}

The Decoder network is a sequential module with an embedding layer, the positional encoding module (same as for the Encoder), a core deep subnetwork, and a posterior distribution prediction layer. The embedding layer transforms a token sequence into a learnable vector representation, and the positional encoding puts temporal information in it. The decoder core subnetwork consists of stacked MHA and FF structures similar to the Encoder but with additional MHA layer between them. Its purpose is to combine the encoder output sequence with the transformed tokens. After this subnetwork completes a step, the last layer yields the next token prediction.

\subsection{Language model}

Language model (LM) predicts the next token and its weight using only the sequence of previous tokens. In the decoding stage, the LM prediction is added to the CTC and Decoder results using a log-linear combination. The model we used is a recurrent neural network of four LSTM layers with 2048 units each.

\subsection{Data preprocessing and augmentation}

For each training set used for experiments, we removed utterances that are too short and too long. After that, the data were speed-perturbed with the perturbation factors 0.9 and 1.1 to make the training set 3 times larger. During the training, acoustic features were augmented with SpecAugment.

\section{TTS system}

We chose a two-network setup, which is prevalent in contemporary neural speech synthesis. The first network (synthesizer) converts input text to a spectrogram. The second network (vocoder) converts input spectrogram to a waveform. The Griffin-Lim algorithm may be used instead of a neural vocoder, but its output usually sounds metallic and less natural.

While ASR takes Mel-spectrograms as an input, converting synthesized spectrograms to waveforms is necessary because of different STFT parameters; unifying them would decrease the quality of either ASR or TTS.

\subsection{Preprocessing}

As LibriSpeech annotations consist of normalized upper-case texts, the only text preprocessing step is G2P. We used lexicon from OpenSLR\footnote{\url{http://www.openslr.org/resources/11/librispeech-lexicon.txt}}. Some of pronunciations in the lexicon were G2P auto-generated.

\subsection{Synthesizer}
\label{synthesizer}

We chose Tacotron \cite{wang2017tacotron, shen2018natural} as a base for our speech synthesizer. It is an RNN-based seq2seq model with attention that converts input text to a log-magnitude 80-band Mel-spectrogram. We used dynamic convolution attention \cite{battenberg2019location} instead of location-sensitive attention \cite{chorowski2015attention} .

We used framework of VAE to model prosody as a deep latent variable. We followed GMVAE-Tacotron \cite{hsu2018hierarchical} and chose prior and posterior distributions to be a mixture of diagonal Gaussians and a single diagonal Gaussian. We used a single global latent variable to model both intra- and inter-speaker prosody variation. Speaker identity and prosody are inherently tied together, so for simplicity we decided to use a single variable to model both.

Loss function is composed of: $l_1$ distance between spectrograms, KL-divergence between prior and posterior of the latent variable \cite[Eq.3]{hsu2018hierarchical}, and CTC loss between spectrogram and input text (following \cite{liu2019maximizing}).

\subsubsection{Architecture details}
Tacotron encoder and decoder hyperparameters follow \cite{battenberg2019effective}: base dimensions are 256 with extensions where concatenation is necessary. For every decoder step, two frames are predicted (r=2). Dynamic convolution attention parameters follow the original paper \cite{battenberg2019location}.

Latent variable encoder consists of two 1D convolutional layers (128 and 256 channels, k=3) with ReLUs and batch normalizations followed by two BLSTM layers (d=128). We applied tanh before projecting to posterior means and log-variances. The latent space is 16-dimensional, the number of components in the prior mixture is 10. Sample from the posterior is concatenated to the encoder output.

Before CTC loss, spectrogram is fed through two linear layers (d=512) with dropout (p=0.5) and ReLU in-between.

\subsection{Vocoder}
Mel-spectrograms produced by Tacotron are smooth and blurry, while the same spectrograms calculated using the output of a neural vocoder are more detailed. We hypothesize that using a vocoder might be beneficial for the final ASR performance.

\begin{figure}[ht]
  \graphicspath{ {./images/} }
  \includegraphics[width=\linewidth]{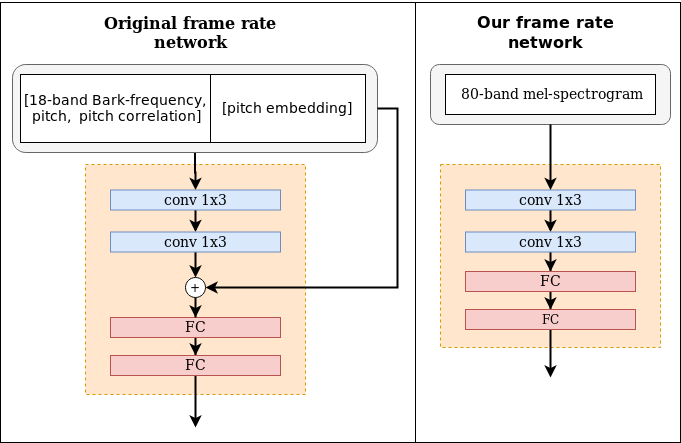}
  \caption{Our LPCNet modification. }
  \label{fig:vocoder}
\end{figure}

Our vocoder is based on LPCNet. It was chosen as a good compromise between quality and inference speed for a multi-speaker TTS system \cite{govalkar2019comparison}.

We modified it to take Mel-spectrograms as an input, as opposed to originally used bark-frequency cepstrum (Figure~\ref{fig:vocoder}). Original LPCNet uses pitch correlation to control sampling temperature, requiring pitch as an input feature. To free Tacotron from predicting pitch and simplify the pipeline, we approximate pitch correlation through spectral flatness. This modification has been successfully used in prior work \cite{Korostik2019TheST}.

The size of the main GRU hidden state was set to 384. We did not apply sparsification. We used the data preparation pipeline from reference implementation with simplifications stated above.

\section{Experiments}

\subsection{ASR setup}
\label{asr_setup}

We took only \textit{train-clean-100} (a hundred-hour portion of the ``clean'' speech) and \textit{train-clean-360} (the rest of the ``clean'' data) for the training, \textit{dev-clean} for tuning of the models, and \textit{test-clean} and \textit{test-other} for the evaluation.  Language model was trained on external text data\footnote{\url{http://www.openslr.org/resources/11/librispeech-lm-norm.txt.gz}}. The acoustic features are cepstral mean and variance normalized 80-dimensional log-Mel filterbank coefficients with 3-dimensional pitch features (fbank-pitch). The text data were tokenized by SentencePiece byte-pair-encoding \cite{kudo_sentencepiece:_2018} with 5000 vocabulary size.

We mostly used the Transformer architecture setup from \cite{karita_comparative_2019} for training\footnote{\url{https://github.com/espnet/espnet/blob/master/egs/librispeech/asr1/conf/tuning/train_pytorch_transformer_large_ngpu4.yaml}} and decoding\footnote{\url{https://github.com/espnet/espnet/blob/master/egs/librispeech/asr1/conf/tuning/decode_pytorch_transformer_large.yaml}}, but our Transformers were trained for 48 epochs with early stopping after five epochs of non-best accuracy for the development set. Also, the parameters averaging was performed over five best-on-devset models, and the beam size was set to 20.

\subsection{TTS setup}

We used 80-band Mel-spectrogram calculated with window of 50 ms and hop of 12.5 ms as an intermediate acoustic representation between synthesizer and vocoder.

\subsubsection{Speech synthesizer}
We trained our TTS on force-aligned \textit{train-clean-100} subset. Input vocabulary consists of English phonemes, a pause break token, and a stop token.

The augmentation process was as follows: we synthesized every text from \textit{train-clean-360} with prosody vector being sampled from the prior. We used one sample per text due to time and computing constraints. We also did not predict pause breaks, which resulted in non-stop continuous speech even for long utterances.

Modeling both speaking style and speaker identity as a single vector imitates a setup where speaker labels are not available (although in a real low-resource setup number of per-speaker data is likely to be more imbalanced). We also believe that \textit{train-clean-100} subset contains enough speakers for the model to generalize; this assumption is yet to be rigorously evaluated.

\subsubsection{Vocoder}
Our LPCNet was trained on \textit{train-clean-100} subset for 250k steps, with a batch size of 64 and with each training sequence consisting of 1600 samples (100ms frame).

We found that for very low-pitched or high-pitched samples the quality of speech signal decreased.

\subsection{Experimental setup}

To establish our low-to-medium-resource baseline, we trained our ASR model on \textit{train-clean-100}. Next, we augmented the data with the speech synthesized from \textit{train-clean-360} texts and trained our TTS-augmented model (\textit{tts-aug-360}). The semi-supervised model was trained on waveforms from the same \textit{train-clean-360} set with transcripts produced by the baseline model (\textit{semi-sup-360}). The inference of the transcripts was performed with the LM according to Section~\ref{asr_setup}. To compare our technique with with the supervised ``oracle'' setup, we trained the model on\textit{train-clean-100} combined with \textit{train-clean-360} (\textit{train-clean-460}).

\subsection{Results}

First of all, we compared our end-to-end system with the best hybrid Kaldi baseline\footnote{\url{https://github.com/kaldi-asr/kaldi/blob/master/egs/librispeech/s5/RESULTS}}. Since our system employs sequential information (via attention decoding mechanism), we compared it with 4-gram LM decoding of the DNN-HMM inference. As shown in Table~\ref{tab:hybrid-vs-e2e}, hybrid system is clearly superior to ours in both \textit{clean} and \textit{other} conditions for \textit{train-clean-100}, while for \textit{train-clean-460} the end-to-end model is better in \textit{clean} and \textit{other} conditions. We also put results for \textit{train-clean-100} from RETURNN \cite{luscher_rwth_2019} as the strongest hybrid baseline known to us.

\begin{table}[ht]
  \caption{Comparison of our end-to-end (E2E) ASR system with Kaldi and RETURNN hybrids on LibriSpeech train-clean-100 and train-clean-460.}
  \label{tab:hybrid-vs-e2e}
  \begin{tabularx}{\linewidth}{@{}|c|c|Y|Y|Y|Y|@{}}
    \hline
    \multirow{3}{*}{Training set} & \multirow{3}{*}{ASR system} & \multicolumn{4}{c|}{WER[\%]}\\
    \cline{3-6}
    & &  \multicolumn{2}{c|}{dev} & \multicolumn{2}{c|}{test}\\
    \cline{3-6}
    & & clean & other & clean & other\\
    \hline
    \hline
    \multirow{3}{*}{\textit{clean-100}} & Kaldi & 5.9 & 20.4 & 6.6 & 22.5\\
    \cline{2-6}
    & RETURNN & \textbf{5.0} & \textbf{19.5} & \textbf{5.8} & \textbf{18.6}\\
    \cline{2-6}
    & E2E (our) & 10.3 & 24.0 & 11.2 & 24.9\\
    \hline
    \hline
    \multirow{2}{*}{\textit{clean-460}} & Kaldi & 5.3 & 17.7 & 5.8 & 19.1\\
    \cline{2-6}
    & E2E (our) & \textbf{4.5} & \textbf{14.1} & \textbf{5.1} & \textbf{14.1}\\
    \hline
  \end{tabularx}
\end{table}

After establishing baselines, we evaluated different ways of converting TTS spectrograms to waveforms. The question was: when synthesized utterances are used for ASR training, are there benefits of choosing neural vocoder over the Griffin-Lim algorithm in terms of final ASR performance? A comparison of Griffin-Lim algorithm with our modification of LPCNet is presented in Table~\ref{tab:GL-vs-our}. As the proposed vocoder delivered 5-6\% relative WER improvement, all synthesized utterances for further experiments were made using the neural vocoder.

\begin{table}[ht]
  \caption{Comparison of waveform synthesis methods. Both ASR models are trained on train-clean-100 plus tts-aug-360.}
  \label{tab:GL-vs-our}
  \begin{tabularx}{\linewidth}{@{}|c|Y|Y|Y|Y|@{}}
    \hline
    \multirow{3}{*}{Vocoder} & \multicolumn{4}{c|}{WER[\%]}\\
    \cline{2-5}
    &  \multicolumn{2}{c|}{dev} & \multicolumn{2}{c|}{test}\\
    \cline{2-5}
    & clean & other & clean & other\\
    \hline
    \hline
    Griffin-Lim & 6.6 & 20.8 & 7.2 & 21.2\\
    \hline
    LPCNet & \textbf{6.3} & \textbf{19.8} & \textbf{6.8} & \textbf{19.9}\\
    \hline
  \end{tabularx}
\end{table}

We thoroughly studied the influence of TTS-augmentation versus semi-supervised learning on the system's performance. Table~\ref{tab:semisup-vs-ttsaug-vs-sup} contains an extensive study on how these approaches perform depending on whether our external LM was used in decoding. Additionally, there is a ``medium-resource'' setup in which \textit{tts-aug-360} is combined with \textit{train-clean-460} instead of \textit{train-clean-100}. The goal of this experiment was to check whether the synthesized speech variability is wide enough to effectively use it with the source utterances of the same text.

The two augmentation approaches without LM usage performed equally and 33\% worse than the supervised one for \textit{test-clean} (lines 2-4). Using decoding with the language model for the TTS-augmented model shortened this performance gap to 23\%, resulting in 4.3 against 3.5 absolute WER respectively (lines 7-9). Compared with the semi-supervised approach, \textit{tts-aug-360} improved WER by 13\% better. Although for \textit{test-other} \textit{semi-sup-360} setup worked better regardless of LM usage. Considering LM influence itself, it had roughly the same effect for the low-to-medium-resource setup (line 6) as the augmentation techniques. Overall, our medium-resource setup benefited from using TTS-augmented data, especially for \textit{test-clean} when LM was employed. Results for \textit{dev-clean} and \textit{dev-other} sets are consistent with the test ones, taking into account a small (about 10\%) over-fitting for \textit{dev-clean}.

\begin{table}[ht]
  \caption{Comparison of different ASR training setups against LM usage.}
  \label{tab:semisup-vs-ttsaug-vs-sup}
  \begin{tabularx}{\linewidth}{@{}|c|c|c|Y|Y|Y|Y|@{}}
    \hline
    \multirow{3}{*}{LM} & \multirow{3}{*}{Core data} & \multirow{3}{*}{Additional data} & \multicolumn{4}{c|}{WER[\%]}\\
    \cline{4-7}
    & & & \multicolumn{2}{c|}{dev} & \multicolumn{2}{c|}{test}\\
    \cline{4-7}
    & & & clean & other & clean & other\\
    \hline
    \hline
    \multirow{5}{*}{No} & \multirow{3}{*}{\textit{clean-100}} & - & 10.3 & 24.0 & 11.2 & 24.9\\
    \cline{3-7}
    & & \textit{tts-aug-360} & 6.3 & 19.8 & \textbf{6.8} & 19.9\\
    \cline{3-7}
    & & \textit{semi-sup-360} & \textbf{6.2} & \textbf{16.9} & \textbf{6.8} & \textbf{17.1}\\
    \cline{2-7} 
    & \multirow{2}{*}{\textit{clean-460}} & - & 4.5 & 14.1 & 5.1 & 14.1\\
    \cline{3-7}
    & & \textit{tts-aug-360} & 4.3 & 13.8 & 4.8 & 13.5\\
    \hline
    \hline
    \multirow{5}{*}{Yes} & \multirow{3}{*}{\textit{clean-100}} & - & 5.8 & 16.6 & 7.0 & 17.0\\
    \cline{3-7}
    & & \textit{tts-aug-360} & \textbf{3.8} & 13.2 & \textbf{4.3} & 13.5\\
    \cline{3-7}
    & & \textit{semi-sup-360} & 4.6 & \textbf{12.7} & 5.2 & \textbf{13.0}\\
    \cline{2-7} 
    & \multirow{2}{*}{\textit{clean-460}} & - & 2.9 & 9.1 & 3.5 & 9.1\\
    \cline{3-7}
    & & \textit{tts-aug-360} & 2.8 & 9.0 & 3.2 & 9.1\\
    \hline
  \end{tabularx}
\end{table}

Difference in improvements on \textit{test-clean} and \textit{test-other} subsets may be attributed to data partition made by LibriSpeech authors. They used ASR performance as the main partition criterion, with ASR being trained on a speech corpus of mostly North American English. As a result, \textit{clean} subset should be, on average, closer to NA accent, and \textit{other} should be farther. Using augmented speech adds bias towards NA accent into the model. Thus, the system is expected to be less effective on non-NA test data due to the domain shift.

The reason for a uneven performance increase when the LM is used might be that the TTS-augmented model's predictions were less confident than produced by the semi-supervised-trained model. This follows from the fact that there are more not-similar-to-anything evaluation data for the \textit{tts-aug-360} setup model than for the \textit{semi-sup-360} one. Thus, we suppose that using the language model, it was easier to shift token probabilities in the right direction.

\begin{table}[ht]
  \caption{Comparison of our system performance against the results of other works for different simulated setups. ``Impr'' stands for relative WER improvement.}
  \label{tab:our-vs-other}
  \begin{tabularx}{\linewidth}{@{}|c|c|Y|Y|Y|Y|@{}}
    \hline
    \multirow{3}{*}{Setup} & \multirow{3}{*}{Paper} & \multicolumn{2}{c|}{WER[\%]} & \multicolumn{2}{c|}{Impr[\%]}\\
    \cline{3-6}
    & & test \newline clean & test \newline other & test \newline clean & test \newline other\\
    \hline
    \hline
    \multirow{3}{*}{low-to-medium-resource} & our & \textbf{4.3} & \textbf{13.5} & \textbf{38.6} & \textbf{20.6}\\
    \cline{2-6}
    & \cite{rosenberg_speech_2019} & 9.3 & 30.6 & 22.8 & 10.1\\
    \cline{2-6}
    & \cite{rossenbach_generating_2020} & 5.4 & 22.2 & 33.3 & 9.4\\
    \hline
    \hline
    \multirow{2}{*}{medium-resource} & our & \textbf{3.2} & \textbf{9.1} & \textbf{8.6} & 0.0\\
    \cline{2-6}
    & \cite{rosenberg_speech_2019} & 6.3 & 22.5 & 0.3 & -0.5\\
    \hline
    \hline
    \multirow{3}{*}{large-resource} & \cite{li2018training} & 4.7 & 15.5 & \textbf{8.6} & \textbf{4.6}\\
    \cline{2-6}
    & \cite{rosenberg_speech_2019} & 4.6 & 13.6 & 4.6 & 1.8\\
    \cline{2-6}
    & \cite{rossenbach_generating_2020} & \textbf{2.5} & \textbf{7.2} & 4.9 & 2.4\\
    \hline
  \end{tabularx}
\end{table}

For low-to-medium- and medium-resource setups, we compared our best results with ones from papers mentioned in Section~\ref{rel_work}. We also provided a large-resource results overview for prior works, although we were unable to successfully train a model on a comparable amount of data due to a lack of computing resources. Note that the partition into resource tasks may not be accurate due to not exactly matching data setups. In Table~\ref{tab:our-vs-other} we considered the final WER and the relative WER improvement from non-augmented setup as two main indicators of good system performance. To save the table space, the results for \textit{dev-clean} and \textit{dev-other} are omitted here. Our best system outperformed previous works for both \textit{test-clean} and \textit{test-other} in the low-to-medium-resource. Despite our system was superior in the medium-resource setup in WER, the improvement diminished for \textit{test-clean} and disappeared for \textit{test-other}. This trend of decreasing improvement is consistent with the results of prior works in the large-resource setup.

\subsection{Limitations}

The data augmentation approach presented in this study has one main limitation -- training data amount and quality. In a conventional low-resource task there may not be enough suitable (with little to no background noise and a diverse enough) audio data to build a working multi-speaker TTS system. In addition, it may be difficult or even impossible for many low-resource languages to find textual data in an amount comparable to that used in this study for LM training.

%Another limitation is speaker prosody variation of synthetic data. Use of the prosody modeling technique described in Section~\ref{synthesizer} may not provide the desired variety, as was observed for non-NA accented \textit{test-other} data.

\section{Conclusion}

In this work we investigated data augmentation for speech recognition using text-to-speech. A low-to-medium-resource setup was simulated with 100-hour subset of LibriSpeech. Using GMVAE-Tacotron as a speech synthesizer and modified LPCNet as a vocoder, we generated 360 hours of synthetic speech with random prosody, modelled by a variational autoencoder. Adding sythesized speech allowed us to improve a powerful end-to-end ASR baseline by 39\% relative WER on \textit{test-clean} and by 21\% on \textit{test-other}. Our approach outperformed similar setups in both absolute WER and relative WER improvement and established a competitive LibriSpeech \textit{train-clean-100} result with 4.3\% WER on \textit{test-clean} and 13.5\% WER on \textit{test-other}. Our experiments also showed that usage of TTS augmentation was more successful than semi-supervised learning on \textit{test-clean}, and less successful on \textit{test-other}.

In future work we plan to address the accent domain shift to improve performance on \textit{test-other} and close the gap to \textit{test-clean} on both low-to-medium- and large-resource tasks. We also plan to evaluate our approach on non-simulated low-resource setups.
% Another study that can be done is to compare MOS with ASR quality to find out whether they are correlated.

\section{Acknowledgements}

This work was partially financially supported by the Government of the Russian Federation (Grant 08-08).



%\bibliographystyle{IEEEtran}

\bibliographystyle{IEEEtran}
\bibliography{mybib}

\begin{thebibliography}{10}
\providecommand{\url}[1]{#1}
\csname url@samestyle\endcsname
\providecommand{\newblock}{\relax}
\providecommand{\bibinfo}[2]{#2}
\providecommand{\BIBentrySTDinterwordspacing}{\spaceskip=0pt\relax}
\providecommand{\BIBentryALTinterwordstretchfactor}{4}
\providecommand{\BIBentryALTinterwordspacing}{\spaceskip=\fontdimen2\font plus
\BIBentryALTinterwordstretchfactor\fontdimen3\font minus
  \fontdimen4\font\relax}
\providecommand{\BIBforeignlanguage}[2]{{%
\expandafter\ifx\csname l@#1\endcsname\relax
\typeout{** WARNING: IEEEtran.bst: No hyphenation pattern has been}%
\typeout{** loaded for the language `#1'. Using the pattern for}%
\typeout{** the default language instead.}%
\else
\language=\csname l@#1\endcsname
\fi
#2}}
\providecommand{\BIBdecl}{\relax}
\BIBdecl

\bibitem{luscher_rwth_2019}
C.~Lüscher, E.~Beck, K.~Irie, M.~Kitza \emph{et~al.}, ``{RWTH} {ASR} {Systems}
  for {LibriSpeech}: {Hybrid} vs {Attention},'' in \emph{Interspeech}, Sep.
  2019, pp. 231--235.

\bibitem{Andrusenko2020TowardsAC}
A.~Andrusenko, A.~Laptev, and I.~Medennikov, ``Towards a competitive end-to-end
  speech recognition for chime-6 dinner party transcription,'' \emph{arXiv
  preprint arXiv:2004.10799}, 2020.

\bibitem{Medennikov_chime6}
I.~Medennikov, M.~Korenevsky, T.~Prisyach, Y.~Khokhlov \emph{et~al.}, ``The stc
  system for the chime-6 challenge,'' in \emph{{CHiME} 2020 {Workshop} on
  {Speech} {Processing} in {Everyday} {Environments}}, 2020.

\bibitem{Wiesner2019}
M.~Wiesner, A.~Renduchintala, S.~Watanabe, C.~Liu \emph{et~al.}, ``{Pretraining
  by Backtranslation for End-to-End ASR in Low-Resource Settings},'' in
  \emph{Interspeech}, 2019, pp. 4375--4379.

\bibitem{synnaeve2019endtoend}
G.~Synnaeve, Q.~Xu, J.~Kahn, E.~Grave \emph{et~al.}, ``End-to-end asr: from
  supervised to semi-supervised learning with modern architectures,''
  \emph{arXiv preprint arXiv:1911.08460}, 2019.

\bibitem{drugman2016active}
T.~Drugman, J.~Pylkkönen, and R.~Kneser, ``Active and semi-supervised learning
  in asr: Benefits on the acoustic and language models,'' in
  \emph{Interspeech}, Sep. 2016, pp. 2318--2322.

\bibitem{shen2018natural}
J.~Shen, R.~Pang, R.~J. Weiss, M.~Schuster \emph{et~al.}, ``Natural tts
  synthesis by conditioning wavenet on mel spectrogram predictions,'' in
  \emph{IEEE International Conference on Acoustics, Speech and Signal
  Processing (ICASSP)}.\hskip 1em plus 0.5em minus 0.4em\relax IEEE, 2018, pp.
  4779--4783.

\bibitem{li2018training}
J.~Li, R.~Gadde, B.~Ginsburg, and V.~Lavrukhin, ``Training neural speech
  recognition systems with synthetic speech augmentation,'' \emph{arXiv
  preprint arXiv:1811.00707}, 2018.

\bibitem{rosenberg_speech_2019}
A.~Rosenberg, Y.~Zhang, B.~Ramabhadran, Y.~Jia \emph{et~al.}, ``Speech
  {Recognition} with {Augmented} {Synthesized} {Speech},'' in \emph{{IEEE}
  {Automatic} {Speech} {Recognition} and {Understanding} {Workshop} ({ASRU})},
  SG, Singapore, Dec. 2019, pp. 996--1002.

\bibitem{tjandra_end--end_2019}
A.~Tjandra, S.~Sakti, and S.~Nakamura, ``End-to-end {Feedback} {Loss} in
  {Speech} {Chain} {Framework} via {Straight}-through {Estimator},'' in
  \emph{{IEEE} {International} {Conference} on {Acoustics}, {Speech} and
  {Signal} {Processing} ({ICASSP})}, Brighton, United Kingdom, May 2019, pp.
  6281--6285.

\bibitem{Ko_2015}
T.~Ko, V.~Peddinti, D.~Povey, and S.~Khudanpur, ``Audio augmentation for speech
  recognition.'' in \emph{Interspeech}, 2015, pp. 3586--3589.

\bibitem{Park_2019}
D.~S. Park, W.~Chan, Y.~Zhang, C.-C. Chiu \emph{et~al.}, ``Specaugment: A
  simple data augmentation method for automatic speech recognition,''
  \emph{Interspeech}, Sep 2019.

\bibitem{rossenbach_generating_2020}
N.~Rossenbach, A.~Zeyer, R.~Schluter, and H.~Ney, ``Generating {Synthetic}
  {Audio} {Data} for {Attention}-{Based} {Speech} {Recognition} {Systems},'' in
  \emph{{IEEE} {International} {Conference} on {Acoustics}, {Speech} and
  {Signal} {Processing} ({ICASSP})}, Barcelona, Spain, May 2020, pp.
  7069--7073.

\bibitem{panayotov_librispeech:_2015-1}
V.~Panayotov, G.~Chen, D.~Povey, and S.~Khudanpur, ``Librispeech: {An} {ASR}
  corpus based on public domain audio books,'' in \emph{{IEEE} {International}
  {Conference} on {Acoustics}, {Speech} and {Signal} {Processing} ({ICASSP})},
  South Brisbane, Queensland, Australia, Apr. 2015, pp. 5206--5210.

\bibitem{griffin_speech_1984}
D.~Griffin, D.~Deadrick, and {Jae Lim}, ``Speech synthesis from short-time
  {Fourier} transform magnitude and its application to speech processing,'' in
  \emph{{IEEE} {International} {Conference} on {Acoustics}, {Speech}, and
  {Signal} {Processing} ({ICASSP})}, vol.~9, San Diego, CA, USA, 1984, pp.
  61--64.

\bibitem{valin2019lpcnet}
J.-M. Valin and J.~Skoglund, ``Lpcnet: Improving neural speech synthesis
  through linear prediction,'' in \emph{IEEE International Conference on
  Acoustics, Speech and Signal Processing (ICASSP)}.\hskip 1em plus 0.5em minus
  0.4em\relax IEEE, 2019, pp. 5891--5895.

\bibitem{yuxuan_style_2018}
Y.~Wang, D.~Stanton, Y.~Zhang, R.~Skerry-Ryan \emph{et~al.}, ``Style tokens:
  Unsupervised style modeling, control and transfer in end-to-end speech
  synthesis,'' in \emph{International Conference on Machine Learning ({ICML})},
  vol.~80, Jul. 2018, pp. 5167--5176.

\bibitem{kingma2013auto}
D.~P. Kingma and M.~Welling, ``Auto-encoding variational bayes,'' in
  \emph{International Conference on Learning Representations (ICLR)}, Dec 2014.

\bibitem{sun2020generating}
G.~Sun, Y.~Zhang, R.~J. Weiss, Y.~Cao \emph{et~al.}, ``Generating diverse and
  natural text-to-speech samples using a quantized fine-grained vae and
  autoregressive prosody prior,'' in \emph{IEEE International Conference on
  Acoustics, Speech and Signal Processing (ICASSP)}.\hskip 1em plus 0.5em minus
  0.4em\relax IEEE, 2020, pp. 6699--6703.

\bibitem{watanabe_espnet:_2018}
S.~Watanabe, T.~Hori, S.~Karita, T.~Hayashi \emph{et~al.}, ``{ESPnet}:
  {End}-to-{End} {Speech} {Processing} {Toolkit},'' in \emph{Interspeech}, Sep.
  2018, pp. 2207--2211.

\bibitem{karita_comparative_2019}
S.~Karita, X.~Wang, S.~Watanabe, T.~Yoshimura \emph{et~al.}, ``A {Comparative}
  {Study} on {Transformer} vs {RNN} in {Speech} {Applications},'' in
  \emph{{IEEE} {Automatic} {Speech} {Recognition} and {Understanding}
  {Workshop} ({ASRU})}, SG, Singapore, Dec. 2019, pp. 449--456.

\bibitem{attention_2017}
A.~Vaswani, N.~Shazeer, N.~Parmar, J.~Uszkoreit \emph{et~al.}, ``Attention is
  all you need,'' in \emph{Advances in Neural Information Processing Systems
  30}, 2017, pp. 5998--6008.

\bibitem{graves_connectionist_2006}
A.~Graves, S.~Fernández, F.~Gomez, and J.~Schmidhuber, ``Connectionist
  temporal classification: labelling unsegmented sequence data with recurrent
  neural networks,'' in \emph{International conference on {Machine} learning -
  {ICML}}, Pittsburgh, Pennsylvania, 2006, pp. 369--376.

\bibitem{wang2017tacotron}
Y.~Wang, R.~Skerry-Ryan, D.~Stanton, Y.~Wu \emph{et~al.}, ``Tacotron: Towards
  end-to-end speech synthesis,'' \emph{arXiv preprint arXiv:1703.10135}, 2017.

\bibitem{battenberg2019location}
E.~{Battenberg}, R.~{Skerry-Ryan}, S.~{Mariooryad}, D.~{Stanton} \emph{et~al.},
  ``Location-relative attention mechanisms for robust long-form speech
  synthesis,'' in \emph{IEEE International Conference on Acoustics, Speech and
  Signal Processing (ICASSP)}, 2020, pp. 6194--6198.

\bibitem{chorowski2015attention}
J.~K. Chorowski, D.~Bahdanau, D.~Serdyuk, K.~Cho, and Y.~Bengio,
  ``Attention-based models for speech recognition,'' in \emph{Advances in
  neural information processing systems}, 2015, pp. 577--585.

\bibitem{hsu2018hierarchical}
W.-N. Hsu, Y.~Zhang, R.~J. Weiss, H.~Zen \emph{et~al.}, ``Hierarchical
  generative modeling for controllable speech synthesis,'' in
  \emph{International Conference on Learning Representations (ICLR)}, 2019.

\bibitem{liu2019maximizing}
P.~Liu, X.~Wu, S.~Kang, G.~Li \emph{et~al.}, ``Maximizing mutual information
  for tacotron,'' \emph{arXiv preprint arXiv:1909.01145}, 2019.

\bibitem{battenberg2019effective}
E.~Battenberg, S.~Mariooryad, D.~Stanton, R.~Skerry-Ryan \emph{et~al.},
  ``Effective use of variational embedding capacity in expressive end-to-end
  speech synthesis,'' \emph{arXiv preprint arXiv:1906.03402}, 2019.

\bibitem{govalkar2019comparison}
P.~Govalkar, J.~Fischer, F.~Zalkow, and C.~Dittmar, ``A comparison of recent
  neural vocoders for speech signal reconstruction,'' in \emph{Proc. 10th ISCA
  Speech Synthesis Workshop}, 2019, pp. 7--12.

\bibitem{Korostik2019TheST}
R.~Korostik, A.~Chirkovskiy, A.~Svischev, I.~Kalinovskiy, and A.~Talanov, ``The
  stc text-to-speech system for blizzard challenge 2019,'' in \emph{Proceedings
  of Blizzard Challenge 2019}, 2019.

\bibitem{kudo_sentencepiece:_2018}
T.~Kudo and J.~Richardson, ``{SentencePiece}: {A} simple and language
  independent subword tokenizer and detokenizer for {Neural} {Text}
  {Processing},'' in \emph{{Conference} on {Empirical} {Methods} in {Natural}
  {Language} {Processing}: {System} {Demonstrations}}, Brussels, Belgium, 2018,
  pp. 66--71.

\end{thebibliography}

\end{document}